# Efficient Bitrate Ladder Construction using Transfer Learning and Spatio-Temporal Features

Ali Falahati[1], Mohammad Karim Safavi[1], Ardavan Elahi[1], Farhad Pakdaman[2], and Moncef Gabbouj[2]

[1] *Avanco*, Tehran, Iran
[2] *Faculty of Information Technology and Communication Sciences, Tampere University*, Tampere, Finland

*Abstract*— Providing high-quality video with efficient bitrate is a main challenge in video industry. The traditional one-size-fits-all scheme for bitrate ladders is inefficient and reaching the best content-aware decision computationally impractical due to extensive encodings required. To mitigate this, we propose a bitrate and complexity efficient bitrate ladder prediction method using transfer learning and spatio-temporal features. We propose: (1) using feature maps from well-known pre-trained DNNs to predict rate-quality behavior with limited training data; and (2) improving highest quality rung efficiency by predicting minimum bitrate for top quality and using it for the top rung. The method tested on 102 video scenes demonstrates 94.1% reduction in complexity versus brute-force at 1.71% BD-Rate expense. Additionally, transfer learning was thoroughly studied through four networks and ablation studies.

## I. Introduction

The vast use of video streaming services has posed challenges for technology providers in maintaining quality of experience (QoE) across devices and networks. Researchers have invested in providing reliable technologies for acquiring, compressing, transmitting, and playing videos. A key aspect of high QoE is smooth playback under varying network conditions. However, compressing video for optimized bitrate transmission while preserving high quality is challenging, as it directly impacts the user's viewing experience.

HTTP Adaptive Streaming (HAS) has become widely used for internet video delivery by storing multiple encoded representations (resolutions and bitrates) of each video on servers. This maximizes quality of experience (QoE) across devices and networks. Traditional "one-size-fits-all" encoding uses the same bitrate ladder for all videos, often wasting bitrate or causing artifacts [1]. Content-adaptive encoding assigns more bitrate to higher complexity videos by encoding multiple times, extracting rate-quality (RQ) curves, and selecting optimal operating points [1]. Some methods first split videos into scenes and apply this process per scene [2], [3]. However, the large number of encodings required for the exhaustive search makes this process computationally complex and impractical with modern codecs [4], requiring optimization.

Recently, machine learning has been used to reduce the computational cost of content-aware encoding [5]–[9]. These methods extract hand-crafted frame-level spatio-temporal features from source videos and employ a machine-learned model to predict parameters used to construct bitrate ladders or to optimize compression [10]. While deep neural networks (DNNs) have shown great performance in computer vision tasks [11], they have not been thoroughly investigated for improved bitrate ladder prediction. DNNs often require huge training datasets to generalize well [12], which can be costly. To address this challenge, transfer learning methods can be employed. These methods utilize the feature maps of a pre-trained network for similar prediction tasks [13]. This idea was recently adopted for video quality assessment to deal with limited datasets [14]–[16].

Another deficiency of existing solutions is that they waste bitrate on the highest bitrate point of the ladder, as they do not consider a proper upper bitrate limit based on each content. The Human Visual System perceives quality as discrete levels [17]. At high bitrates, quality degradation cannot be perceived until a threshold. Studies show VMAF scores above 95 (or 92) cannot be discriminated from reference video [18][19]. Ignoring this wastes bitrate. To remedy this, we introduce the Highest Quality (HQ) point as the minimum bitrate with unperceived degradations at the top ladder rung.

Previous works studied correlations between low-level spatial/temporal features or encoder-related features, and RQ behavior. However, using transfer learning and pre-trained networks for bitrate ladder prediction has not been thoroughly investigated. Considering this, we investigate this relationship and propose a content-aware prediction scheme employing feature maps extracted from pretrained DNNs. The architecture consists of pretrained DNN modules to extract spatio-temporal features. Features are combined and processed with Gated Recurrent Units (GRUs) to capture temporal dependencies. Using transfer learning, pretrained models are adapted for ladder construction, i.e., to predict parameters of key ladder operating points. Next, pre-encodings are performed to capture Rate-Quality characteristics at the key points to construct the ladder. Inspired by [9], we demonstrate correlations between RQ characteristics of adjacent resolutions, reducing pre-encodings required. Transfer learning is thoroughly studied using four pretrained models and ablation studies. The contributions of this paper are as follows:

- A near-optimal bitrate ladder construction method using pre-trained DNNs, leading to more accurate prediction and significantly reducing pre-encodes for construction.

- A thorough study of transfer learning for ladder construction, testing four pre-trained networks and ablation studies.

- Propose a method to model and use an HQ point for the highest rung of the bitrate ladder, corresponding to minimum bitrate for highest quality, saving bitrate for high quality scenarios compared to existing solutions.
- A large dataset of spatio-temporal features for all scenes, facilitating further research. Scenes were separated manually from sequences for accurate detection. The dataset and codes will be released on https://github.com/researchSME/dnn-ladder-predictor.

The rest of the paper is organized as follows: Section II details the DNN architecture and modeling procedure. Section III describes steps to predict and construct a near-optimal bitrate ladder. Section IV explains constructing the reference ladder. Section V discusses the experimental results, and Section VI concludes the paper.

## II. PREDICTION MODELS USING TRANSFER LEARNING

Fig. 1 describes the overall workflow of the proposed approach in both training and testing phases. First, the prediction models are trained in a supervised manner using the extracted deep features from pre-trained DNNs and specific RQ points as prediction target. Then, the trained models are used to predict the RQ points and to construct the predicted bitrate ladder. While Section III describes the ladder construction steps, this section describes the proposed model, including network architecture and training details.

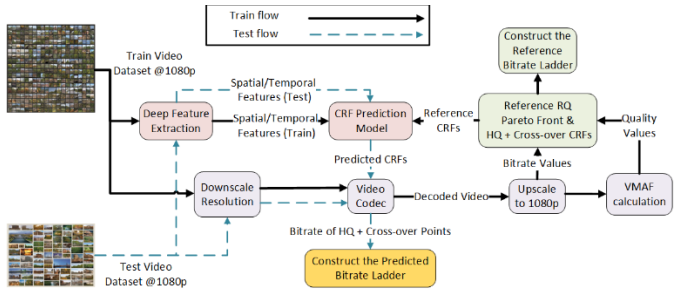

Fig. 1. Overall workflow of the paper

### A. Prediction Model Architecture

Fig. 2 details the proposed network architecture, which is inspired by the network used in [15] for quality assessment. The network includes pre-trained spatial and temporal feature branches, as well as learnable components to adapt to the task at hand.

**Deep Spatial Feature Extractors**: Image classification models pre-trained on large datasets like ImageNet [12] contain rich content information. Transfer learning is commonly used to benefit from these models for alternative tasks, removing the need for large training data/resource. To this end we selected the following image classification models and their combinations for spatial feature extraction: (1) ResNet50 [20], (2) VGG16 [21], and (3) InceptionV3 [22]. All models are pre-trained on ImageNet.

**Deep Temporal Feature Extractor:** In addition to spatial complexity, video's temporal complexity plays a great role in its compressibility. To incorporate motion information, we use a computationally light pre-trained 3D-CNN - specifically, the fast pathway of the SlowFast model [23] trained on Kinetics-400 [24] for temporal feature extraction.

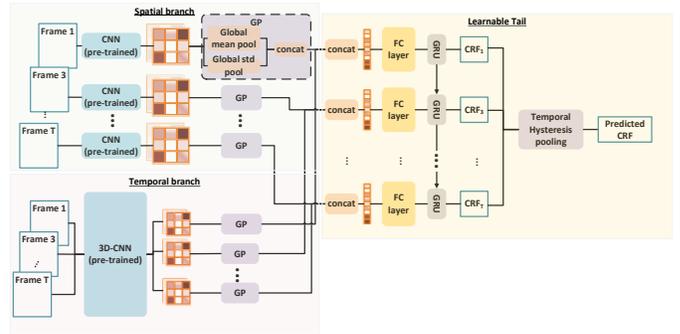

Fig. 2. Proposed model architecture

**Predicting a Target RQ Point:** To predict desired points on the RQ curve, $T$ frames from each uncompressed video scene are passed to pretrained models. Frames are selected from the center of the scene, and if the scene is shorter than $T$ frames, all frames are used. The feature maps from the last layer of each model are extracted, and global average and standard deviation pooling are applied to reduce the feature size. The output feature size of each model with respect to $T$ is listed in Table I. We observed that $T = 240$ frames are enough to capture the spatial and temporal information from the scenes. The SlowFast model produces $T/2$ feature maps, as it produces one feature map for every two frames. The feature vectors of all models are concatenated to form the combined feature vector. If the model feature sizes differ in the first dimension, the features are down sampled to match the one with the lowest length. A GRU [25] is used to model temporal dependencies among frames of a video sequence, similar to [26]. Before feeding the per-frame features into the GRU, a fully connected (FC) layer is used to reduce their size. The sequence of GRU hidden states is then converted into a score value for each frame using an FC layer. The per-frame scores are aggregated into a single score for the entire video sequence using the temporal hysteresis pooling method inspired by the human visual system (HVS) [15]. In contrast to [15], where the score values are mapped to subjective video quality scores, we map them to specific Constant Rate Factor (CRF) values on the RQ curve of the video scene. Since the deep features are extracted from well-trained models, only a small dataset is needed to train the rest of the architecture.

TABLE I. OUTPUT DIMENSION OF MODELS FOR $T$ FRAMES

| Pre-trained model | Output feature size |
|---|---|
| ResNet-50 | $T \times 4096$ |
| VGG16 | $T \times 1024$ |
| InceptionV3 | $T \times 1024$ |
| SlowFast | $T/2 \times 512$ |

### B. Training Configuration

To perform the supervised training of the described architecture, we used Adaptive Moment Estimation (Adam [27]) optimizer which is widely used for training deep models. The loss function used for training consists of three terms. The first term is a Pearson Linear Correlation Coefficient (PLCC) loss, inspired by [15], which optimizes the model for prediction precision. The second term is a Spearman Rank-order

Correlation Coefficient (SRCC) loss, which encourages the monotonicity of the predictions. Additionally, an L1 loss between the actual and predicted points is included to minimize prediction error. The final loss function is as follows:

$$l = l_{PLCC} + \lambda l_{SRCC} + \gamma l_{l1} \quad (1)$$

Where $\lambda$ and $\gamma$ can be tuned to find the best converging model. By changing these parameters, we observed that $\lambda = \gamma = 1$ is a good balance between the losses. The configuration parameters that we used are listed in Table II. We use the described model to construct the near-optimal bitrate ladder in the next section.

TABLE II. MODEL CONFIGURATION PARAMETERS

| Parameter | Value |
|---|---|
| FC layer output size | 270 |
| GRU output size | 32 |
| Learning rate | $5e-4$ |
| $\lambda$ | 1 |
| $\gamma$ | 1 |

## III. PREDICTING THE BITRATE LADDER

This section presents the proposed bitrate ladder prediction approach using the model from Section II. The target resolution set is defined as S={1080p,720p,480p,360p}, with a lower bitrate limit of $R_{min}$ =150Kbps. There is no specific high bitrate limit, and the highest predicted bitrate is content-dependent. Libx265 [28], an open-source H.265/HEVC codec, is used for CRF-based encodings. FFmpeg [29] v4.3 is used to perform the encodings. Also, VMAF [30] is used as the objective quality metric to construct the bitrate ladder.

To find the near-optimal ladder, we start with the highest resolution $(S_1)$. The first predicted point is the "Highest Quality" (HQ) point, where quality difference between encoded and uncompressed video is unnoticeable by HVS. This is the point where allocating more bits results in no improvement in quality, and allocating less leads to quality degradation. There have been studies mapping this point to a VMAF score. Authors in [19] show that VMAF=95 provides over-the-top quality, and in most cases, VMAF=92 has only a negligible subjective mean-opinion-score (MOS) difference from the original video. Therefore, we select VMAF=92 as our HQ point to avoid wasting bitrate. The CRF value reaching VMAF=92 at $S_1$ is predicted to find this point.

After predicting the HQ point CRF, we use this point as the first row of the optimal ladder (top rung). To find subsequent rows, we predict resolution cross-over points where optimal resolution switches to next lower one, similar to [9]. These points along with the HQ point are shown in Fig. 3. An independent CRF predictor is trained to predict each of the crossover points, or the HQ point. Having two crossover CRFs ($CRF_{S_i}^{low}$ and $CRF_{S_i}^{high}$) for each resolution, we encode the video at these points to obtain crossover bitrates. Note that for the highest resolution we only have one crossover point ($CRF_{S_1}^{high}$). Instead of $CRF_{S_1}^{low}$ we encode at the HQ point ($CRF_{S_1}^{HQ}$).

As seen in Fig. 4, a strong linear correlation exists between adjacent resolution crossover CRFs ($CRF_{S_i}^{high}$ and $CRF_{S_{i+1}}^{low}$) with a PLCC over 0.97. Hence, to reduce complexity, only $CRF_{S_i}^{low}$ needs predicting while $CRF_{S_i}^{high}$ can be inferred from predicted $CRF_{S_{i+1}}^{low}$. In total, one CRF value is predicted per resolution: $CRF_{pred} = \{CRF_{S_1}^{HQ}, CRF_{S_2}^{low}, CRF_{S_3}^{low}, CRF_{S_4}^{low}\}$ and the rest are inferred.

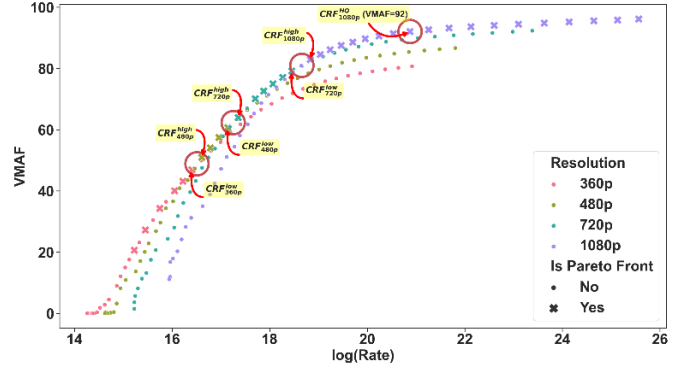

Fig. 3. VMAF-bitrate curve example from "Meridian" scene in different resolutions. The points marked by "X" are on the Pareto Front.

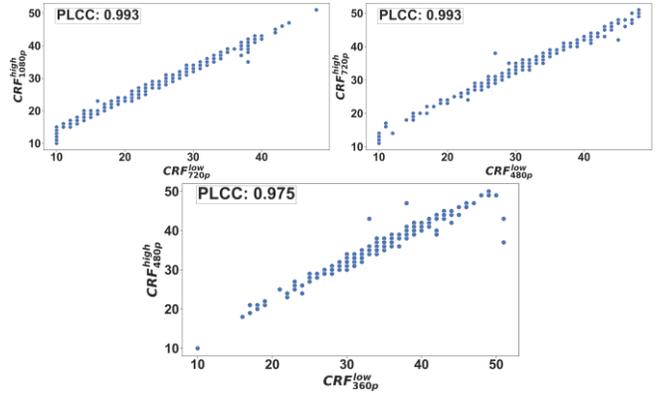

Fig. 4. Scatter plots of adjacent cross-over CRF pairs.

We explored the relation between CRF and $\log_2(Rate)$ of compressed videos, observing a strong linear relation with an $R^2$ of around 0.99. An example for the "Bosphorus" video scene can be observed in Fig. 5(a). Therefore, to determine the nearest CRF corresponding to a target bitrate at each resolution, we use the following equation:

$$CRF_s = \zeta_S \log_2(Rate) + \delta_S \quad (2)$$

Where $\zeta_S$ and $\delta_S \in R$ for each resolution $S$ are determined from the HQ and crossover CRF-resolution points already encoded. Moreover, to further reduce the number of pre-encodes, the relationship between $\zeta_S$ and $\delta_S$ of each resolutions was studied. $\zeta_{360p}$ was found to have a strong linear relationship with $\zeta_{480p}$, as seen in Fig. 5(b). This means only one CRF-rate pair needs encoding at in 360p instead of two. These findings agree with similar findings in [9] reported for QP value.

By predicting the HQ point CRF, crossover CRFs per resolution, and deriving $\zeta_S$ and $\delta_S$ per resolution, enough information is obtained to construct a close-to-optimal bitrate ladder. A common rule to construct the final optimal ladder is to use an arbitrary bitrate step rule between rows. (3) is used where $R_i$ is the bitrate of the $i$-th row and $K$ is a constant between 1.5-2. A lower $K$ results in more rows/storage, while a higher $K$ reduces rows but deteriorates quality. Based on number of encodes for different $K$ values, $K=2$ was determined as the proper value in our work.

$$R_{i-1} \cong K * R_i \quad (3)$$

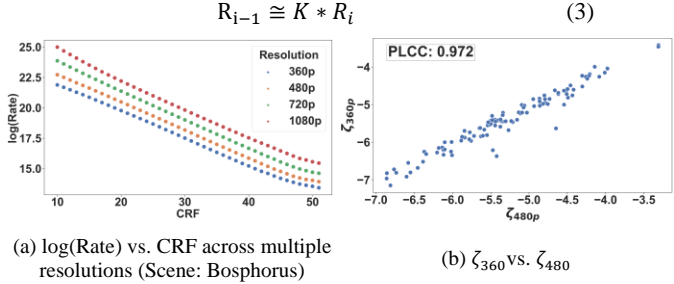

(a) log(Rate) vs. CRF across multiple resolutions (Scene: Bosphorus)

(b) $\zeta_{360}$ vs. $\zeta_{480}$

Fig. 5. Examining bitrate vs. CRF parameters.

## IV. REFERENCE BITRATE LADDER CONSTRUCTION

This section explores the 2D rate-quality space across resolutions, defines the Pareto front (PF) in this space, and how ground truth HQ and crossover points are obtained for training the CRF predictors. It also discusses the construction of the reference bitrate ladder (RL), which serves as the ground truth for evaluation.

### A. Description of the Dataset

A large and comprehensive dataset of over 680 video scenes from multiple open-source datasets (UVG [31], LIVE-APV [32], AVT-VQDB-UHD-1 [33], Xiph.org (derf [34] and av2 [35] collections), Tears of Steel [36], SVT [37], and SJTU [38]) was gathered. All sequences were downscaled to 1920x1080 and converted to 4:2:0 chroma subsampling for consistency. Frame rates were 24, 30 or 60 fps. Each sequence consists of a single video scene and is at least 30 frames long. Scenes were manually detected to eliminate automatic scene detection errors. 70% of the data were used for training, 15% for testing, and 15% for validation. The scenes represent a wide range of spatial and temporal complexity as seen in Fig. 6. In this figure Spatial Information (SI) and Temporal Information (TI) [39] are used to show the content variety of the videos in the dataset.

### B. Constructing the Reference Pareto Front

Finding an optimal bitrate ladder is a multi-objective optimization problem with bitrate and quality as the two optimization objectives. These problems have a set of optimal solutions, the Pareto Front (PF). To create the reference PF for the video sequences we first encode each sequence using all CRF values in the [10,51] range at each resolution in $S$. A total of 168 encodes are performed per sequence. Then VMAF is calculated between the decompressed and the uncompressed videos to create RQ points. To construct the PF from these operating points, the convex hull of the points is extracted. The top-left surface of the convex-hull is used as the PF. After the PF is calculated, the points in each resolution after which a resolution switch occurs are extracted as the cross-over points and the point in $S_1$ resolution with VMAF score closest to 92 is considered as the HQ point. An example of the PF and ground truth points can be observed in Fig. 3.

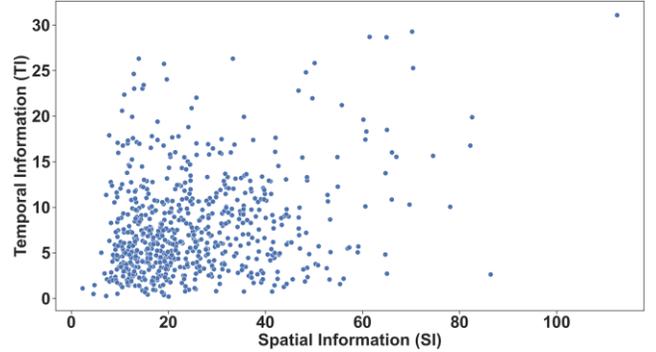

Fig. 6. SI and TI distribution of the dataset.

### C. Constructing the Reference Bitrate Ladder

To construct the final reference ladder from the PF, we first add the CRF-resolution pair of the HQ point to the bitrate ladder. Then the bitrate of this point is divided by the constant $K$ from (3) and the CRF-resolution pair of the point on the PF with bitrate closest to this point is appended to the ladder. We repeat this step until the bitrate of the resulting encode is lower than the minimum bitrate $R_{min}$.

## V. EXPERIMENTAL RESULTS

This section evaluates the proposed method from multiple aspects. First, the performance of the CRF predictor is evaluated. Here the feature extractors of Section II.A are utilized in different combinations as shown in Table III. The feature combination with best performance per computational cost is selected and its final performance is assessed using Bjøntegaard Delta (BD) metrics. Next, the impact of predicting the HQ point is assessed. Finally, an ablation study is carried out. The implementation was done using the PyTorch framework.

### A. Choosing the Spatial Feature Extractor

To find the best spatial feature extractor, CRF prediction accuracy of different feature combinations is compared in Table III. In all cases temporal features are concatenated to the spatial features. ResNet-50 achieves lowest BD-rate vs RL but InceptionV3 is fastest to extract features. Combining ResNet-50 with VGG-16 improves accuracy but with added complexity. ResNet-50 provides a good tradeoff between accuracy and computation cost based on these results. Even when not best, mean-absolute-error (MAE) difference of ResNet-50 from best model is small, implying enough information is gathered without extra computation. This results also indicate that the CRF prediction model proposed in Section II performs better predicting HQ point CRF vs crossover CRFs.

### B. BD Metrics and Computational Complexity

Fig. 7 shows the distributions of BD metrics vs RL for the test dataset using ResNet50+SlowFast features. A BD-rate and BD-VMAF closer to zero indicates a predicted ladder closer to its corresponding RL. The average BD-rate and BD-VMAF loss of the proposed method for the 102 videos are 1.71% and -0.26 respectively, requiring significantly fewer pre-encodes and VMAF calculations than alternative methods. The proposed

TABLE III. PREDICTION PERFORMANCE OF DIFFERENT SPATIAL FEATURE COMBINATIONS

| Features* | BD-rate(%) vs. RL | Inference Time (s) | $CRF_{S_1}^{HQ}$ | | | | $CRF_{S_2}^{low}$ | | | | $CRF_{S_3}^{low}$ | | | | $CRF_{S_4}^{low}$ | | | |
|---|---|---|---|---|---|---|---|---|---|---|---|---|---|---|---|---|---|---|
| | | | MAE | $\Delta VMAF < 6$ | $\Delta CRF < 2$ | SROCC | MAE | $\Delta VMAF < 6$ | $\Delta CRF < 2$ | SROCC | MAE | $\Delta VMAF < 6$ | $\Delta CRF < 2$ | SROCC | MAE | $\Delta VMAF < 6$ | $\Delta CRF < 2$ | SROCC |
| **ResNet-50** | 1.71 (±2.23) | 54.80 | 1.869 | 94.23% | 77.89% | 0.774 | 4.561 | 62.50% | 45.19% | 0.745 | 3.546 | 38.00% | 45.00% | 0.761 | 3.049 | 35.21% | 50.70% | 0.659 |
| VGG16 | 1.98 (±2.52) | 79.32 | 1.954 | 93.27% | 72.11% | 0.784 | 5.070 | 54.81% | 37.50% | 0.601 | 3.696 | 44.00% | 49.00% | 0.773 | 3.077 | 36.62% | 49.30% | 0.627 |
| InceptionV3 | 3.30 (±3.75) | 39.95 | 2.806 | 82.69% | 56.73% | 0.690 | 5.936 | 50.00% | 36.54% | 0.528 | 4.638 | 33.00% | 38.00% | 0.724 | 3.794 | 28.17% | 42.25% | 0.455 |
| ResNet50 + VGG16 | 1.85 (±2.20) | 131.34 | 1.856 | 94.23% | 78.85% | 0.799 | 4.743 | 61.54% | 43.27% | 0.758 | 3.623 | 38.00% | 45.00% | 0.761 | 3.108 | 35.21% | 46.48% | 0.609 |
| Resnet50 + InceptionV3 | 2.90 (±3.31) | 91.97 | 2.676 | 85.58% | 57.69% | 0.751 | 5.456 | 54.81% | 40.38% | 0.506 | 4.552 | 32.00% | 41.00% | 0.688 | 3.626 | 29.58% | 40.84% | 0.478 |
| VGG16 + InceptionV3 | 3.30 (±3.72) | 116.49 | 2.482 | 89.42% | 61.54% | 0.741 | 5.656 | 50.96% | 37.50% | 0.634 | 4.870 | 31.00% | 38.00% | 0.701 | 3.841 | 25.35% | 40.84% | 0.513 |
| ResNet-50 + VGG16 + InceptionV3 | 3.29 (±3.61) | 168.51 | 2.816 | 84.62% | 56.73% | 0.726 | 5.676 | 55.77% | 37.50% | 0.526 | 4.723 | 31.00% | 37.00% | 0.663 | 3.796 | 29.58% | 40.84% | 0.547 |

*Features from the SlowFast model are included in all rows.

method requires $2|S| - 1 = 7$ pre-encodes. The number of encodes required after the pre-encodes to construct the ladder varies for each scene, averaging 9.83 encodes per video sequence over the test set, a 94.1% reduction compared to brute force method.

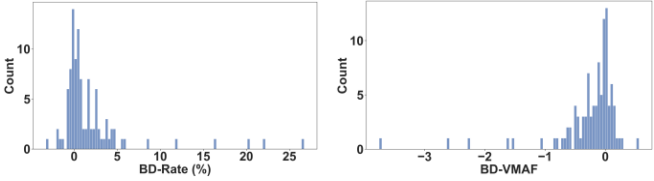

Fig. 7. BD-rate and BD-VMAF distribution of predicted bitrate ladders compared to ground truth. A BD-rate and BD-VMAF closer to zero indicates a predicted ladder closer to the ground truth.

Fig. 8 shows predicted ladders alongside their corresponding RL for several test videos. The first four show successful prediction cases by the proposed method. The last row depicts unsuccessful cases. For "TunnelFlag", the HQ point predictor fails, leading to an unnecessarily high bitrate at the top rung. For "controlled_burn", the crossover predictors fails, resulting in a high BD-rate. This demonstrates both successful and less successful prediction examples.

In Table IV, the results are compared to the Feature-based Ladder prediction (FL) method from [9] using PSNR-based BD-rate vs RL. FL employs handcrafted features for prediction, while the proposed method uses CNN features. As observed in the table, the results are close but the proposed method achieves better outcomes in some cases and on average. The number of encodes required for the two methods is similar as they follow a similar approach.

TABLE IV. RESULTS COMPARISON WITH [9]

| Sequence | BD-rate (%) | |
|---|---|---|
| | Ours | [9] |
| Boxing Practice | 1.36 | 0.54 |
| Coastguard | 4.19 | 0.03 |
| Crosswalk | 0.42 | 0.59 |
| Treeshade | 1.29 | 1.61 |
| WindAndNature-scene2. | 0.94 | 10.21 |
| Average | 1.64 | 2.60 |

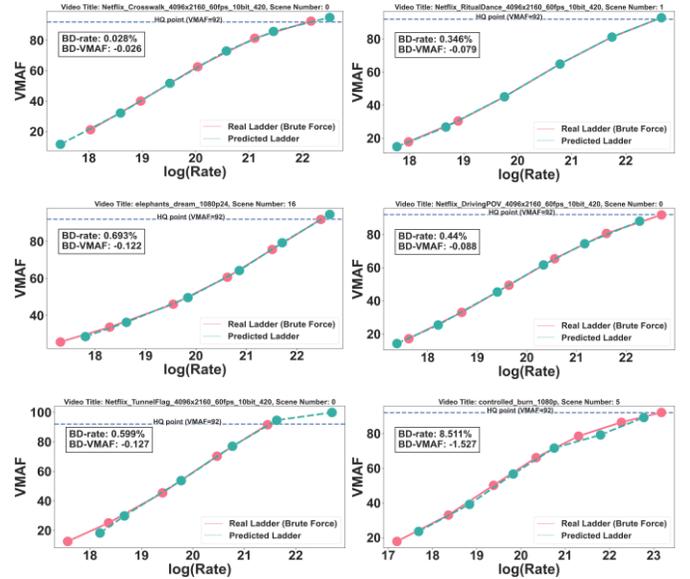

Fig. 8. Predicted Vs. Reference bitrate ladder examples.

In Table V, the proposed method is compared to [40] in terms of BD-rate vs RL and total number of encodes (including pre-encodes). The comparison uses shared video scenes in both papers' test sets. The proposed method requires significantly fewer encodes to construct the ladder, though the average BD-rate is higher. However, per-sequence BD-rates are close in most cases. Given features come from unrelated pre-trained models, this is still a significant result for the proposed method.

### C. Impact of Predicting the HQ Point

To evaluate the impact of HQ point, the ladder construction was repeated without predicting the HQ point. Instead of using $CRF_{S_1}^{HQ}$, $CRF_{S_1}^{low} - J$ was used as the first rung CRF, where $J = 5$ provided the best value for this constant. The VMAF difference of the first row from 92 ($\Delta VMAF = VMAF_{first\ row} - 92$) was measured for all test sequences. This allowed comparison to the proposed method that does predict the HQ point.

TABLE VII. ABLATION STUDY RESULTS

| Case | BD-rate(%) vs. RL | Inference Time (s) | $CRF_{S_1}^{HQ}$ | | | | $CRF_{S_2}^{low}$ | | | | $CRF_{S_3}^{low}$ | | | | $CRF_{S_4}^{low}$ | | | |
|---|---|---|---|---|---|---|---|---|---|---|---|---|---|---|---|---|---|---|
| | | | MAE | $\Delta VMAF < 6$ | $\Delta CRF < 2$ | SROCC | MAE | $\Delta VMAF < 6$ | $\Delta CRF < 2$ | SROCC | MAE | $\Delta VMAF < 6$ | $\Delta CRF < 2$ | SROCC | MAE | $\Delta VMAF < 6$ | $\Delta CRF < 2$ | SROCC |
| **ResNet-50+Slowfast (with GRU)** | 1.71 (±2.23) | 54.80 | 1.869 | 94.23% | 77.89% | 0.774 | 4.561 | 62.50% | 45.19% | 0.745 | 3.546 | 38.00% | 45.00% | 0.761 | 3.049 | 35.21% | 50.70% | 0.659 |
| ResNet-50 only (with GRU) | 1.97 (±2.33) | 52.02 | 2.296 | 91.35% | 62.5% | 0.576 | 5.145 | 56.73% | 39.42% | 0.663 | 3.723 | 38.00% | 45.00% | 0.739 | 3.473 | 29.58% | 43.66% | 0.520 |
| Slowfast only (with GRU) | 2.05 (±2.68) | 2.78 | 2.064 | 94.23% | 68.27% | 0.736 | 5.340 | 55.77% | 31.73% | 0.488 | 4.407 | 40.00% | 46.00% | 0.534 | 3.693 | 28.17% | 40.84% | 0.436 |
| ResNet-50+Slowfast (no GRU) | 1.89 (±2.28) | 54.80 | 1.917 | 91.35% | 76.92% | 0.757 | 4.680 | 63.46% | 44.23% | 0.700 | 3.659 | 40.00% | 48.00% | 0.763 | 3.392 | 22.54% | 39.44% | 0.586 |
| ResNet-50 only (no GRU) | 1.84 (±2.31) | 52.02 | 2.354 | 90.38% | 63.46% | 0.511 | 5.007 | 57.69% | 42.31% | 0.682 | 3.904 | 40.00% | 48.00% | 0.732 | 3.531 | 26.76% | 36.62% | 0.524 |
| Slowfast only (no GRU) | 2.70 (±3.09) | 2.78 | 2.208 | 90.38% | 68.27% | 0.72 | 5.660 | 50.00% | 35.58% | 0.421 | 4.777 | 31.00% | 38.00% | 0.541 | 3.754 | 28.17% | 39.44% | 0.488 |

TABLE V. RESULTS COMPARISON WITH [40]

| Sequence | BD-rate (%) | | # of encodes | |
|---|---|---|---|---|
| | Ours | [40] | Ours | [40] |
| Touchdown pass | 0.56 | 0.12 | 10 | 20 |
| Pedestrian area | -0.66 | 0.00 | 11 | 22 |
| Blue sky | 2.9 | 0.26 | 11 | 18 |
| Runners | 0.03 | 0.05 | 11 | 18 |
| Netflix Tango | 0.66 | 0.00 | 11 | 21 |
| Netflix Ritual Dance | 0.24 | 0.00 | 11 | 18 |
| Netflix Crosswalk | 0.03 | -0.23 | 12 | 22 |
| Netflix Bar Scene | 5.45 | 0.10 | 8 | 20 |
| Netflix Driving POV | -0.15 | -0.84 | 11 | 15 |
| Fountains | 1.79 | 0.02 | 13 | 13 |
| GTA V scene | 1.63 | 0.92 | 12 | 18 |
| Beauty | 2.46 | 1.12 | 11 | 23 |
| Average | 1.25 | 0.13 | 11 | 19 |

Fig. 9 shows the distribution of ΔVMAF differences with and without HQ point prediction. Predicting HQ point results in a more concentrated distribution around zero, indicating more efficient coding. Two scenarios are considered: 1) $VMAF_{first\ row} > 92$ wastes bitrate without quality gain 2) $VMAF_{first\ row} < 92$ loses quality and hurts the user experience. Without HQ point, quality varies significantly. Table VI summarizes results for both cases, with $\Delta Rate = Rate_{first\ row} - Rate_{VMAF=92}$ and $\Delta VMAF = VMAF_{first\ row} - 92$. Without HQ point, on average 3.03 Mbps of bitrate is wasted (when predicting the first row with higher VMAFs), and 4.6 VMAF points of quality is lost (when predicting the first row with a low VMAF). When HQ is used, these values are 1.19 Mbps and 2.88 VMAF points, respectively.

TABLE VI. EFFECT ANALYSIS FOR PREDICTING THE HQ POINT

| Method | $VMAF_{first\ row} > 92$ | | $VMAF_{first\ row} < 92$ | |
|---|---|---|---|---|
| | $\Delta Rate$ (Mbps) | $\Delta VMAF$ | $\Delta Rate$ (Mbps) | $\Delta VMAF$ |
| with HQ | 1.19(±0.92) | 3.58(±2.00) | -3.70(±4.16) | -2.88(±1.68) |
| w/o HQ | 3.03(±2.45) | 4.91(±2.19) | -4.28(±4.58) | -4.60(±3.29) |

*D. Ablation Study*

An ablation study was conducted to measure the impact of different parts of the bitrate ladder prediction process. The impact of spatial and temporal features was studied by measuring prediction performance with and without these features. To validate GRU's effectiveness, the process was also repeated without GRU. Table VII summarizes the results. By observing MAE of predictions, removing either temporal or spatial features from feature extraction significantly degrades performance. Similarly, removing GRU from CRF prediction degrades performance but to a lesser extent. This validates the importance of each component of the proposed model.

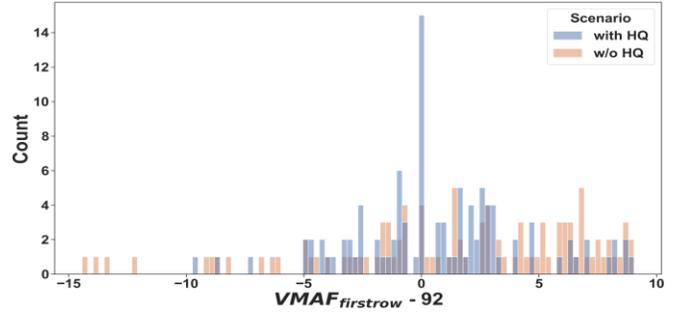

Fig. 9. Histogram of the VMAF difference between the first row of the predicted ladder and 92, with and without predicting the HQ point.

## VI. CONCLUSION

This work proposed an efficient per-scene bitrate ladder construction method using transfer learning with spatial and temporal features from pre-trained models. Predicting the minimum bitrate point (HQ point) to achieve excellent visual quality for the top rung was proposed to comply with HVS. Moreover, the use of transfer learning was thoroughly investigated, by experimenting with four different pre-trained networks, and via an ablation study. Results demonstrate that (1) using DNNs via transfer learning can improve performance over existing methods, (2) Using the HQ point optimizes bitrate use by guaranteeing high quality and avoiding waste, (3) Both spatial and temporal network branches are vital for high performance. However, adding complexity by combining features from multiple CNNs is unnecessary, and most pre-trained networks are sufficient for feature extraction if adapted properly for the ladder construction task.

ACKNOWLEDGEMENT

This project received funding from the European Union's Horizon 2020 research and innovation programme under the Marie Skłodowska-Curie grant agreement No [101022466]. The authors also wish to acknowledge Avanco, for supporting this research with computational resources.

REFERENCES

[1] "Per-Title Encode Optimization. delivering the same or better… , by Netflix Technology Blog , Netflix TechBlog." https://netflixtechblog.com/per-title-encode-optimization-7e99442b62a2 (accessed May 20, 2023).

[2] "High Quality Video Encoding at Scale , by Netflix Technology Blog , Netflix TechBlog." https://netflixtechblog.com/high-quality-video-encoding-at-scale-d159db052746 (accessed May 20, 2023).

[3] "Dynamic optimizer — a perceptual video encoding optimization framework , by Netflix Technology Blog , Netflix TechBlog." https://netflixtechblog.com/dynamic-optimizer-a-perceptual-video-encoding-optimization-framework-e19f1e3a277f (accessed May 20, 2023).

[4] F. Pakdaman, M. A. Adelimanesh, M. Gabbouj, and M. R. Hashemi, "Complexity Analysis of Next-Generation VVC Encoding and Decoding," in *Proceedings - International Conference on Image Processing, ICIP*, 2020, pp. 3134–3138.

[5] J. Šuljug and S. Rimac-Drlje, "Fast Content Adaptive Representation Selection in HEVC-Based Video Coding for Streaming Applications," in *Proceedings Elmar - International Symposium Electronics in Marine*, 2022, pp. 169–174.

[6] M. Bhat, J. M. Thiesse, and P. Le Callet, "A case study of machine learning classifiers for real-time adaptive resolution prediction in video coding," in *Proceedings - IEEE International Conference on Multimedia and Expo*, 2020, pp. 1–6.

[7] V. V. Menon, H. Amirpour, M. Ghanbari, and C. Timmerer, "Perceptually-Aware Per-Title Encoding for Adaptive Video Streaming," in *Proceedings - IEEE International Conference on Multimedia and Expo*, 2022, pp. 1–6.

[8] F. Nasiri, W. Hamidouche, L. Morin, N. Dholland, and J. Y. Aubie, "Ensemble Learning for Efficient VVC Bitrate Ladder Prediction," *Proc. - Eur. Work. Vis. Inf. Process. EUVIP*, vol. 2022-Septe, 2022, [Online]. Available: http://arxiv.org/abs/2207.10317

[9] A. V. Katsenou, J. Sole, and D. Bull, "Efficient Bitrate Ladder Construction for Content-Optimized Adaptive Video Streaming," *IEEE Open J. Signal Process.*, vol. 2, pp. 496–511, 2021.

[10] A. Elahi, A. Falahati, F. Pakdaman, M. Modarressi, and M. Gabbouj, "NCOD: Near-Optimum Video Compression for Object Detection," in *IEEE International Symposium on Circuits and Systems (ISCAS)*, 2023, pp. 1–5.

[11] A. Krizhevsky, I. Sutskever, and G. E. Hinton, "ImageNet classification with deep convolutional neural networks," *Commun. ACM*, vol. 60, no. 6, pp. 84–90, 2017.

[12] Jia Deng, Wei Dong, R. Socher, Li-Jia Li, Kai Li, and Li Fei-Fei, "ImageNet: A large-scale hierarchical image database," in *2009 IEEE conference on computer vision and pattern recognition*, 2009, pp. 248–255.

[13] S. Kornblith, J. Shlens, and Q. V. Le, "Do better imagenet models transfer better?," in *Proceedings of the IEEE Computer Society Conference on Computer Vision and Pattern Recognition*, 2019, pp. 2656–2666.

[14] D. Li, T. Jiang, and M. Jiang, "Quality assessment of in-the-wild videos," in *Proceedings of the 27th ACM International Conference on Multimedia*, 2019, pp. 2351–2359.

[15] B. Li, W. Zhang, M. Tian, G. Zhai, and X. Wang, "Blindly Assess Quality of In-The-Wild Videos via Quality-Aware Pre-Training and Motion Perception," *IEEE Trans. Circuits Syst. Video Technol.*, vol. 32, no. 9, pp. 5944–5958, 2022.

[16] W. Sun, X. Min, W. Lu, and G. Zhai, "A Deep Learning based No-reference Quality Assessment Model for UGC Videos," in *MM 2022 - Proceedings of the 30th ACM International Conference on Multimedia*, 2022, pp. 856–865.

[17] T. Tian, H. Wang, L. Zuo, C. C. J. Kuo, and S. Kwong, "Just noticeable difference level prediction for perceptual image compression," *IEEE Trans. Broadcast.*, vol. 66, no. 3, pp. 690–700, 2020.

[18] C. G. Bampis, Z. Li, I. Katsavounidis, T. Y. Huang, C. Ekanadham, and A. C. Bovik, "Towards Perceptually Optimized Adaptive Video Streaming-A Realistic Quality of Experience Database," *IEEE Trans. Image Process.*, vol. 30, no. c, pp. 5182–5197, 2021.

[19] A. Kah, C. Friedrich, T. Rusert, C. Burgmair, W. Ruppel, and M. Narroschke, "Fundamental relationships between subjective quality, user acceptance, and the VMAF metric for a quality-based bit-rate ladder design for over-the-top video streaming services," in *Applications of Digital Image Processing XLIV*, 2021, p. 38.

[20] K. He, X. Zhang, S. Ren, and J. Sun, "Deep residual learning for image recognition," in *Proceedings of the IEEE Computer Society Conference on Computer Vision and Pattern Recognition*, 2016, pp. 770–778.

[21] K. Simonyan and A. Zisserman, "Very deep convolutional networks for large-scale image recognition," *3rd Int. Conf. Learn. Represent. ICLR 2015 - Conf. Track Proc.*, 2015.

[22] C. Szegedy, V. Vanhoucke, S. Ioffe, J. Shlens, and Z. Wojna, "Rethinking the Inception Architecture for Computer Vision," in *Proceedings of the IEEE Computer Society Conference on Computer Vision and Pattern Recognition*, 2016, pp. 2818–2826.

[23] C. Feichtenhofer, H. Fan, J. Malik, and K. He, "Slowfast networks for video recognition," in *Proceedings of the IEEE International Conference on Computer Vision*, 2019, pp. 6201–6210.

[24] W. Kay *et al.*, "The Kinetics Human Action Video Dataset," *arXiv Prepr. arXiv1705.06950*, 2017, [Online]. Available: http://arxiv.org/abs/1705.06950

[25] J. Chung, C. Gulcehre, K. Cho, and Y. Bengio, "Empirical Evaluation of Gated Recurrent Neural Networks on Sequence Modeling," *arXiv Prepr. arXiv1412.3555*, 2014, [Online]. Available: http://arxiv.org/abs/1412.3555

[26] D. Li, T. Jiang, and M. Jiang, "Unified Quality Assessment of in-the-Wild Videos with Mixed Datasets Training," *Int. J. Comput. Vis.*, vol. 129, no. 4, pp. 1238–1257, 2021.

[27] D. P. Kingma and J. Ba, "Adam: A method for stochastic optimization," *arXiv Prepr. arXiv1412.6980*, 2014.

[28] "x265, the free H.265/HEVC encoder - VideoLAN." https://www.videolan.org/developers/x265.html (accessed Sep. 22, 2023).

[29] "FFmpeg." https://ffmpeg.org/ (accessed Dec. 24, 2023).

[30] Zhi Li; Anne Aaron; Ioannis Katsavounidis; Anush Moorthy; Megha Manohara, "Toward A Practical Perceptual Video Quality Metric , by Netflix Technology Blog , Netflix TechBlog," 2016. https://netflixtechblog.com/toward-a-practical-perceptual-video-quality-metric-653f208b9652 (accessed Aug. 23, 2023).

[31] A. Mercat, M. Viitanen, and J. Vanne, "UVG dataset: 50/120fps 4K sequences for video codec analysis and development," in *MMSys 2020 - Proceedings of the 2020 Multimedia Systems Conference*, 2020, pp. 297–302.

[32] Z. Shang, J.P. Ebenezer, Y. Wu, H. Wei, S. Sethuraman, and A. C. Bovik, "LIVE-APV Live Video Streaming Database." https://live.ece.utexas.edu/research/LIVE_APV_Study/apv_index.html (accessed Sep. 21, 2023).

[33] R. R. Ramachandra Rao, S. Goring, W. Robitza, B. Feiten, and A. Raake, "AVT-VQDB-UHD-1: A Large Scale Video Quality Database for UHD-1," in *Proceedings - 2019 IEEE International Symposium on Multimedia, ISM 2019*, 2019, pp. 17–24.

[34] Xiph.org, "Xiph.org :: Derf's Test Media Collection," 2015. http://media.xiph.org/video/derf/ (accessed Sep. 21, 2023).

[35] "Xiph.org :: AV2 Candidate Test Media." https://media.xiph.org/video/av2/ (accessed Sep. 21, 2023).

[36] "Tears of Steel , Mango Open Movie Project." https://mango.blender.org/ (accessed Sep. 21, 2023).

[37] "SVT Open Content Video Test Suite 2022 – Natural Complexity." https://www.svt.se/open/en/content/ (accessed Sep. 21, 2023).

[38] L. Song, X. Tang, W. Zhang, X. Yang, and P. Xia, "The SJTU 4K video sequence dataset," in *2013 5th International Workshop on Quality of Multimedia Experience, QoMEX 2013 - Proceedings*, 2013, pp. 34–35.

[39] ITU-T Recommendation P.910, "Subjective video quality assessment methods for multimedia applications," *Int. Telecommun. Union, Geneva*, pp. 1–42, 2009.

[40] S. Paul, A. Norkin, and A. C. Bovik, "Efficient Per-Shot Convex Hull Prediction By Recurrent Learning," pp. 1–11, 2022, [Online]. Available: http://arxiv.org/abs/2206.04877